\documentclass[sigconf]{acmart}

\AtBeginDocument{%
  \providecommand\BibTeX{{%
    \normalfont B\kern-0.5em{\scshape i\kern-0.25em b}\kern-0.8em\TeX
    
    }}}

\usepackage{xcolor}

\usepackage{soul}

\setstcolor{red}

\copyrightyear{2024}
\acmYear{2024}
\setcopyright{rightsretained}
\acmConference[CHI '24]{Proceedings of the CHI Conference on Human Factors in Computing Systems}{May 11--16, 2024}{Honolulu, HI, USA}
\acmBooktitle{Proceedings of the CHI Conference on Human Factors in Computing Systems (CHI '24), May 11--16, 2024, Honolulu, HI, USA}
\acmDOI{10.1145/3613904.3642452}
\acmISBN{979-8-4007-0330-0/24/05}

\usepackage{multirow}
\begin{document}

\title{Situating Data Sets: Making Public Data Actionable for Housing Justice}

\author{Anh-Ton Tran}
\orcid{0000-0002-5381-7159}
\affiliation{%
  \institution{Georgia Institute of Technology}
  %\streetaddress{North Ave NW}
  \city{Atlanta}
  \state{Georgia}
  \country{USA}
  \postcode{30332}
}
\email{atran91@gatech.edu}

\author{Grace Guo}
\orcid{0000-0001-8733-6268}
\affiliation{%
  \institution{Georgia Institute of Technology}
  %\streetaddress{North Ave NW}
  \city{Atlanta}
  \state{Georgia}
  \country{USA}
  %\postcode{30332}
}
\email{gguo31@gatech.edu}

 \author{Jordan Taylor}
 \orcid{0000-0002-0896-992X}
 \affiliation{
   \institution{Carnegie Mellon University}
  \city{Pittsburgh}
  \state{PA}
 \country{USA}
 }
 \email{jordant@andrew.cmu.edu}

\author{Katsuki Chan}
\orcid{0009-0004-7275-2233}
\affiliation{%
  \institution{Georgia Institute of Technology}
  %\streetaddress{North Ave NW}
  \city{Atlanta}
  \state{Georgia}
  \country{USA}
  %\postcode{30332}
}
\email{kchan90@gatech.edu}

\author{Elora Raymond}
\orcid{0000-0002-9459-561X}
\affiliation{%
  \institution{Georgia Institute of Technology}
  %\streetaddress{North Ave NW}
  \city{Atlanta}
  \state{Georgia}
  \country{USA}
  %\postcode{30332}
}
\email{elora.raymond@design.gatech.edu}

\author{Carl DiSalvo}
\orcid{0000-0002-7344-8741}
\affiliation{%
  \institution{Georgia Institute of Technology}
  %\streetaddress{North Ave NW}
  \city{Atlanta}
  \state{Georgia}
  \country{USA}
  %\postcode{30332}
}
\email{cdisalvo@gatech.edu}

\begin{abstract}
Activists, governments, and academics regularly advocate for more open data. But how is data made open, and for whom is it made useful and usable? In this paper, we investigate and describe the work of making eviction data open to tenant organizers. We do this through an ethnographic description of ongoing work with a local housing activist organization. This work combines observation, direct participation in data work, and creating media artifacts, specifically digital maps. Our interpretation is grounded in D’Ignazio and Klein’s Data Feminism, emphasizing standpoint theory. Through our analysis and discussion, we highlight how shifting positionalities from data intermediaries to data accomplices affects the design of data sets and maps. We provide HCI scholars with three design implications when situating data for grassroots organizers: becoming a domain beginner, striving for data actionability, and evaluating our design artifacts by the social relations they sustain rather than just their technical efficacy.

\end{abstract}




\maketitle

\section{Introduction}

Critical data scholars have often pointed out that data cannot speak for itself \cite{loukissas2019all,d2020data}. Data is produced and used in particular settings and conditions \cite{sambasivan2021everyone,gitelman2013raw,bowker2000sorting}. Despite frequent calls for making civic data sets open, we often do not consider the contingency of this data and the privileges necessary to use ``open'' data. \cite{boyd2012critical,loukissas2019all,d2020data}. This paper details the process of making a particular civic data set -- aggregated court eviction records (court data) -- actionable to grassroots, anti-eviction tenant organizing.

One influential contribution to critical data approaches has been D'ignazio and Klein's \textit{Data Feminism}, which utilizes feminist theory to guide just data practices \cite{d2020data}. 
Since the release of \textit{Data Feminism}, researchers have leveraged arguments from D'Ignazio and Klein's work to evaluate outcomes ranging from algorithmic justice to effective data visualization and communication, and Open Data projects \cite{lee2021viral, hill2020stake,li2022approaching, paudel2023reimagining}. Our paper uses Data Feminism to inform how we conduct engaged work with data, similar to recent research from the Data Feminist lab \cite{suresh2022towards}.
In the context of making civic data public, \textit{Data Feminism} notes how these efforts are often ``raw data dumps'' \cite{d2020data}, couched in formats that are common to us as data scientists and HCI scholars: spreadsheets, CSVs, and JSON files. These formats imply a particular user, one who is versed in working with data \cite{loukissas2019all}. Even though we may refer to public data as ``raw'', these data sets are typically already cleaned and standardized to a degree that further abstracts them from their context \cite{loukissas2019all}.

Our work dives into these critical considerations of open data described by \textit{Data Feminism} through an in-depth design case study working with court data. We conduct this work with a tenant's rights organization based in Atlanta, a large metropolitan area in the U.S. South with one of the highest eviction filing rates in the country \cite{raymondMetro2020}. 
The court data was the product of an institutional effort to scrape the data from local court record portals, and we were tasked with identifying ways to make this data set useful for community-based organizations.
Such a task encompassed understanding the data provenance, translating what this data can achieve for our partners, cleaning and parsing the data, and producing data visualizations.
This process spanned 1.5 years, and our engagement with our community partners is ongoing. In detailing this case, we provide insights into how we put elements of Data Feminism and critical data approaches into practice, especially in contexts -- such as tenant organizing -- that challenge entrenched power structures.

As D'ignazio and Klein emphasize, working with data does not simply result in the development of an artifact. It is a process where the technical expert's work is to scaffold various lines of use and value to the user \cite{d2020data} -- in this case, our community partners. The court data challenged us in how we interpreted it and what we could achieve to meet the needs of tenant organizing. While we developed a data manual and Participatory Design workshop, these methods were not enough to elicit use of the court data. Situating this data set for our community partners required us to re-situate our positionality as technical experts and the values we bring to working with the data. Only after joining our partners in organizing for housing justice could we identify what makes this court data useful in their context.  
The process of shifting our standpoints \cite{haraway1988situated,harding1986science} allowed us to understand the standpoint of our partners, a prerequisite to truly make a data set actionable to a particular setting. Describing this case highlights all of the forms of data work that are underreported and have cascading effects on the civic data interventions we wish to design 
\cite{sambasivan2021everyone}.

We share these findings to contribute to community-engaged research in HCI, particularly work involving data. Our paper expands on previous activist-oriented HCI research that exemplifies critical considerations and reflexivity of our positionalities as those with expertise \cite{whitney2021hci,ghoshal2020toward, halperin2023temporal}.
We leave readers with three implications to apply to their own research. First, shifting positionalities requires a willingness to be a domain novice in their partner's expertise. In our case, becoming a novice at tenant organizing helped us better match our technical expertise to the needs of our partners. Second, we expand on the vision of Open Data by arguing data should not only be available and accessible, but actionable. This entails connecting the context of the data to the situatedness of the actions your partner wishes to take \cite{suchman1987plans}. Finally, we ask HCI scholars to not only consider how our technical interventions serve functional needs, but foster relations. These relations may supercede functionality, which forces us to evaluate our artifacts not by their technical aspects but their social impact.

\section{Related Works} 

\subsection{Data Feminism and Critical Data}
Our work draws heavily from \textit{Data Feminism} by D’Ignazio and Klein. In their book, they connect feminist epistemology with concepts from critical data studies and HCI \cite{d2016feminist}, providing guidelines on stewarding a design process that is inclusive and plural across the various points of engagement. We use Data Feminism to inform two threads of our research, grappling with "open" data and how to structure community data efforts.

The concept of Open Data emerged out of a demand to make public records freely accessible for use and re-distribution \cite{handbook2015open}. The idealized vision of Open Data imagines applications and services built off civic data, enabling greater transparency and capacity for the public to hold governments accountable \cite{d2020data}. These goals reflect the desires of our partners with using public eviction records. However, most Open Data fails to live up to this promise due to a lack of adequate resources, and instead we are left with "raw data dumps" \cite{davies2017exploring}. These formats result in ``zombie data'', left unused since they are not easily accessible to the general public \cite{d2020data,gurin2014open}. D'Ignazio and Klein call attention to this issue in their book, arguing that data scientists should look to feminist theory to provide the context needed to support clear value and use to various potential stakeholders of public data \cite{d2020data}. Other HCI scholars have used Data Feminism to interrogate Open Data. Paudel and Soden, for instance, examine Open Government Data projects for disaster relief efforts using Data Feminism as an evaluation framework. \cite{paudel2023reimagining}. Our research similarly examines public government data, however, we are not using Data Feminism to evaluate others but instead to structure our own design process with a community organization.

In \textit{Data Feminism}, D'Ignazio and Klein outline a set of principles that apply the tenets of feminism and standpoint theory to data work.
These principles helped guide and inform our design process and engagement with our community partners, and served as critical questions we used to interrogate our work. They are as follows:

%HOW TO MAKE A LIST
 \begin{enumerate}
     \item \textbf{Examine Power:} How does power operate in the world? \label{examine_power}
    
     \item \textbf{Challenge Power:} How is this work committed towards challenging unequal power structures to work towards justice? \label{challenge_power}
    
     \item \textbf{Elevate Emotion \& Embodiment:} How does this work recognize multiple forms of knowledge, particularly the recognition that humans are corporeal in this world? \label{elevate}
    
     \item \textbf{Rethink Binaries and Hierarchies:} How do we challenge existing systems of counting and classification that oppress like the gender binary? \label{binaries_and_hierarchies} 
    
    \item \textbf{Embrace Pluralism:} How does this work synthesize multiple perspectives since no single knowledge is complete?\label{embrace_pluralism}
    
     \item \textbf{Consider Context:} How does this work account for the locality and context of data? \label{consider_context}
     
     \item \textbf{Make Labor Visible:} How does this work make the labor of data visible? \label{make_labor_visible}
    \cite{d2020data}
 \end{enumerate}

These principles are distilled from feminist theories into implementable questions and guidelines. However, this translation still leaves much up to the researcher or data scientist to figure out once brought into a project. 
We thus turned to feminist approaches in HCI research that have set out a research agenda coupling feminist principles to methodologies \cite{bardzell2010feminist,bardzell2011towards}, as well as detailed applications of Data Feminism in HCI work. 
% Therefore, we looked at works that put Feminist epistemology into HCI practice\cite{dimond2013hollaback}. 
In this vein, we parallel Suresh et al., who apply Data Feminism in their co-design engagement with activists to develop data sets and machine learning models on femicide \cite{suresh2022towards}. Our work unfolds in a different context -- evictions -- and application -- data visualizations. 

Throughout their book, D'ignazio and Klein weave elements of Feminist Standpoint Theory to explain the importance of positionality in equitably structuring collaborations. Standpoint Theory developed in the 1970s and 1980s as a feminist critique of the practices of knowledge production, challenging the view that politics mar scientific inquiry \cite{harding2004feminist}.
Standpoint scholars argue that all knowledge production is partial and contingent \cite{haraway1988situated}, situated in specific histories and practices shaped "by gender, class, race, and culture" \cite{harding1986science}. Instead of neutrality, standpoint theoriests argue for \textit{strong objectivity} and \textit{situated knowledges}. In HCI, Suchman parallels standpoint theory in regards to user tasks, arguing that actions are situated in particular contexts instead of based on abstractions and generalized principles \cite{suchman1987plans}. Standpoint theory also presents itself as a methodology and political strategy to fight against oppression and empower marginalized groups. Many of these instantiations of standpoint theory are developed out of Black Feminist thought \cite{collins1990black,crenshaw1990mapping}. We turned to standpoint theory to consider our roles as technical experts entering a social context where Black and Communities of Color are overrepresented in systemic marginalization \cite{fields2021racialized}.

\subsubsection{Critical Data Studies}

Critique of data, data systems, and data labor have been a longstanding interest area for HCI and adjacent fields. Many early studies explored the disciplining power of categories in computing \cite{suchman1993categories}, the social construction of classification \cite{bowker2000sorting}, and the politics of how work is represented in CSCW systems \cite{star1999layers}. 
% Similar interest in the ways people make sense of data exists in HCI research on sensemaking, or "the process of searching for a representation and encoding data in that representation to answer task-specific questions" \cite{russell1993cost}. This work on sensemaking has guided research on user interface design and visualization in HCI for decades. More contemporary research on sensemaking has looked at the partial, contingent ways both AI developers \cite{cabrera2023did} and everyday users \cite{shen2021everyday} understand algorithmic behavior. 
% Recent research on data work practices has also investigated how the the design of algorithmic fairness tool-kits envision the AI ethics work \cite{wong2023seeing} and the systemic undervaluing of data labour relative to model development \cite{sambasivan2021everyone}.
In conversation with this research, the more recent interdisciplinary field of critical data studies has developed over the past decade amid growing interest in Big Data and AI. In 2012, boyd \& Crawford described the ways Big Data can obfuscate its social construction, leading to false invocations of objectivity and accuracy \cite{boyd2012critical}. Crawford et al. expand on this by emphasizing ethical, methodological, epistemological, and cultural critiques of Big Data \cite{crawford2014big}. Dalton \& Thatcher then coined the term ``critical data studies", reminding readers that data is situated and calling out Big Data's appearance of neutrality. The overarching argument of these critiques is the acknowledgment that data is neither neutral nor does data ever speak for itself \cite{d2020data,loukissas2019all}. Instead, all data comes from somewhere, and its interpretation requires people \cite{jasanoff2017visible,loukissas2019all}. 
For those learning to see the world through data, Passi \& Jackson emphasize the need to "straddle the competing demands of formal abstraction and empirical contingency" \cite{passi2017data}, and acknowledge how trust and use of data systems is a product of complex collaboration \cite{passi2018trust}. This highlights the importance of considering the situated, partial, and political ways that data represents the world, as well as the complex ways in which people interpret data. Data is shaped by the context in which it is created and the values and decisions of those who create, use, and interpret it.
In this paper, we describe the process and challenges of contextualizing evictions data and putting it to use to support the activities of our community partners.

\subsection{Critical Visualization and Counter-mapping}

\subsubsection{Critical Visualization}
Research on critical visualization intersects with both critical data approaches in HCI and the emerging interdisciplinary field of critical data studies. Like data work, visualization is a discipline that imposes implicit values on how data is represented and communicated. Early visualization studies prioritized insight generation from data sets \cite{card1999readings, fekete2008value, north2006toward, chang2009defining}, valuing sense-making \cite{pirolli2005sensemaking}, and the accuracy and speed of information retrieval \cite{cleveland1984graphical, heer2010crowdsourcing, tufte1985visual, tufte1985visual}.
These pragmatic considerations led to the establishment of data visualization conventions such as preferential encodings \cite{cleveland1984graphical, heer2010crowdsourcing}, minimizing the data-ink ratio \cite{tufte1985visual, tufte1985visual}, Shneiderman's mantra \cite{shneiderman2003eyes}, and the use of dynamic queries \cite{shneiderman1994dynamic, williamson1992dynamic, ahlberg2003visual}.

Many visualization researchers have since questioned these prescriptive metrics and design conventions.
Some have revisited the universal application of best practices that prioritize precision and information retrieval \cite{correll2014bad, bertini2020shouldn}.
Others call for more critical approaches like examining contextual factors \cite{peck2019data}, contending with impact and harm \cite{correll2019ethical}, and grappling with the inherent power dynamics and outcomes engendered \cite{dork2013critical}.
In \textit{Data Feminism}, D'Ignazio and Klein examine and unpack many of these arguments for critical visualization \cite{d2020data}.
Their close reading of situated and community-engaged visualization work shows how existing visualization guidelines must be extended to value contextual factors above conventional prescriptive metrics and design conventions.

These shifts in visualization scholarship have led to an emphasis on \textit{designing with} communities.
Participatory methodologies -- such as the nine-step design study approach proposed by Sedlmair et al. \cite{sedlmair2012design}, the action design research approach \cite{mccurdy2016action}, and the design activity framework \cite{mckenna2014design} -- aim to solve real-world problems by centering user experiences in context.
However, they still reinforce conventional roles where the researcher is the \textit{visualization expert} and the user is the \textit{domain expert}.
Recent work by Akbaba et al. \cite{akbaba2023troubling} analyzes collaborative visualization projects through the lens of care. They advocate for extending existing methodologies, grounding care ``in the complexity of politics, power, and interpersonal relationships'' as a form of situated knowledge-making.
Our work expands on these prior frameworks by detailing the life cycle of a visualization design process where existing methodologies are brought into politically charged, community-based contexts.

\subsubsection{Counter-mapping}
Closely related to critical visualization and critical data is the field of critical cartography and techniques of counter-mapping.
Maps, like visualizations, are situated in complex realities that cannot be abstracted away through simple 2D representations \cite{shelton2022situated}.
Critical cartography challenges the semblance of neutrality and objectivity in conventional cartography research, and interrogates the power of mapping in shaping political interests \cite{harley2002new, crampton2018introduction, kim2015critical}.
In particular, counter-mapping has been developed as a way of subverting cartographic neutrality to highlight the emotional and personal experiences of place-making, challenging power, building public coalitions, and supporting grassroots organizing \cite{peluso1995whose, carrera2023unseen, louis2012introduction, hazen2006power}.
Dalton \& Thatcher explore the history of counter-mapping and the use of maps that “challenge predominant power effects of mapping” and “engages in mapping that upset[s] power relations” \cite{harris2005power, dalton_thatcher_2014}, as inspiration for critical engagement using Big Data.

The challenges of simultaneously building maps for housing justice and politically accompanying our partners in day-to-day organizing work made us question the value of our artifacts and its connection to tenant organizing. Carrera et al. have conducted an interview study with grassroots organizers on their use of counter mapping to support their abolitionist goals \cite{carrera2023unseen}. Their findings helped us orient the labor involved in our collaboration with HJL. Specific to housing justice activism, Shelton grapples with fighting the "gods-eye trick \cite{haraway1988situated}" and leveraging data and objectivity to appeal to the state \cite{shelton2022situated}. These impulses mirror concerns that surface in Carrera's study \cite{carrera2023unseen} and helped us to identify the work of our maps which are situated in specific organizing tasks. Using data on vacant and abandoned properties, Shelton uses a form of situated mapping to demonstrate how maps can serve both pragmatic and critical goals \cite{shelton2022situated}. 
The Anti-Eviction Mapping Project has also taken up the call for counter-mapping by visualizing evictions on maps to highlight the prevalence of displacement and present the oral histories from people impacted by evictions \cite{maharawal2018anti, halperin2023probing, halperin2023temporal}. Our work is heavily inspired by AEMP's projects, especially their critical reflection
on building resource maps for tenant organizing during the Covid-19 Pandemic \cite{halperin2023temporal}. Halperin and McElroy detail the pitfalls and disruptions when technical experts parachute into counter-mapping projects with little context about the dynamic housing and eviction crisis landscape \cite{halperin2023temporal}. This work helped us to reflect on our roles, the values we bring, and how we deploy our expertise with our partners.
In this paper, we build on this corpus of critical cartography and counter-mapping research to explore the design and use of maps to support the everyday tasks of housing activism.

\subsection{Community-Based Research and Housing}
Finally, our work also pulls from community and activist-oriented scholarship in HCI \cite{asad2019prefigurative,whitney2021hci,ghoshal2019role,ghoshal2020toward}. 
Ghoshal's work attends to value misalignment between technology tools and the grassroots organizations that use them, ultimately arguing for developing technologies based on the cultures of the organizations themselves \cite{ghoshal2019role,ghoshal2020toward}. 
Meanwhile, Whitney et al. argue for HCI scholars to engage in other activities besides designing technological tools for a grassroots organization \cite{whitney2021hci}. 
We take that call in earnest through our long-term commitment to our partners. 

Our research context is in housing, specifically evictions. Housing is a growing area of interest for HCI scholars, particularly those committed to community partnerships and social justice.
Corbett and Loukissas call for HCI work that examines and addresses issues of gentrification as a social justice research agenda. They define gentrification as ``a collective process of settlement by higher-income people in a low-income area, resulting in the forced class and race-based displacement of existing residents'' \cite{corbett2019engaging}. Our work is situated within this call for HCI \cite{corbett2019engaging} to address these wicked problems \cite{rittel1974wicked}.
Recently, Halperin et al. designed a conversational agent to probe how AI can benefit storytelling to support grassroots social movements. They interview participants who narrate or collect stories of housing insecurity and eviction \cite{halperin2023probing}. 
Additionally, Asad and LeDantec have investigated how ICT intersects with civic engagement. 
They conducted ethnographic fieldwork with housing activists to understand ways to support non-legitimate forms of civic action in foreclosure blockades \cite{asad2015illegitimate}.
Our work shares the same context of housing activism in the U.S. South, and we build on Asad's contribution by detailing how we open and situate a data set for activists.

We also took note of housing-related HCI research, which focuses on data and grassroots contexts. 
Zegura et al. describe how care practices were employed in their project with a community housing advocacy group to collate civic and grassroots collected data. 
They wanted to understand how to use data science for social good, outside of logics of efficiency \cite{zegura2018care}. 
Meng et al. collaborate with a grassroots organizer to engage in community data collection of neighborhood code violations. 
This data work is described within the framework of a caring democracy to demonstrate how data collection can encourage community members to act for better living conditions \cite{meng2019collaborative}. 
We build on these studies by contributing findings from working with institutional housing data sets in a grassroots context and designing data visualizations to support community organizing.

\section{Background: Eviction Records} \label{background}
This paper details our work with eviction data from the city of Atlanta.  
Since the Covid-19 pandemic began, eviction filings have continued in the U.S. despite policies like the federal eviction moratorium.
The Federal Reserve Bank of Atlanta (FED), a governmental institution, and the Atlanta Regional Commission (ARC), an intergovernmental coordinating agency in charge of regional planning, partnered with our Georgia Tech's School of City and Regional Planning to gather data to understand the state of evictions during the pandemic. 
They built a webscraper to pull eviction records from each of the five metro counties' online public court record portals \cite{raymondMetro2020}. 
We will refer to this data set as the \textit{court data}. 
These institutions are politically limited in what they could do with this data due to institutional requirements for accuracy, precision, and a high burden of proof. 
How they communicate findings from the court data must
maintain a sense of objectivity to mitigate reputational risk, ensuring the broader public trusts these institutions.  
Most of the artifacts produced by these organizations take the form of public briefings and reports \cite{raymond2021special}. These formats are useful to policymakers and public officials who employ the data findings towards writing legislation.  
We draw a parallel between this approach to presenting data to Haraway’s framing of the objective gaze and standpoint theory's critique of science \cite{haraway1988situated, harding1986science}.
While these findings and insights published in the briefings can be valuable, this take on the data is situated and partial despite being couched in seemingly objective formats.
Resources published by these institutions may not be useful in the same way to a community organization.

In contrast, grassroots organizations are not beholden to this level of proof. 
A grassroots organization may be more willing to stake a claim with data to center the concerns of tenants who face an unfair housing system.  
Whereas large institutions want to seem unbiased, grassroots organizations may take an explicitly political stance to express a clear viewpoint. 
This difference in what an organization can express with data is not lost on the institutions.
% Our work to design these visualizations can be framed as a form of “institutioning” -- participatory design within institutions. Intermediation is a form of institutioning where institutions identify ways to adopt PD processes in their own contexts \cite{teli2020tales}.
The FED and ARC could make the court data ``open'' and ``public'', but they could not use it directly towards mobilizing communities. 
However, they could hope for others to find uses for the data by community organizations, which is why one of the data architects brought us on as
data intermediaries (DIs).
% This was the impetus behind our involvement in this data set by one of the data architects.
DIs are actors and groups that bridge the gap between public government data and groups that require scaffolding of technical expertise to access and use such data \cite{meng2019social, sein2011between, chattapadhyay2014access, dombrowski2014government}. These tasks include compiling, formatting, cleaning, and designing the data \cite{chattapadhyay2014access}.

The organization we designed these visualizations with is HJL, a tenants’ rights organization. They support tenants to self-organize and form unions to build collective power against landlords. As DIs, we could facilitate access to this data while providing expertise to parse and make sense of it.
Through this work, the first author became a full-time member and organizer at HJL, which involved attending bi-weekly meetings and operational support for direct action mobilization. 
%Membership is voluntary, and the organization is open to anyone interested in sharing their commitments to housing justice. 
The organization takes a consensus-based decision-making process, distributing organizing work across members within a working group.
The second author joined the first as a visualization expert, attending meetings during the artifacts' design, development, and implementation. 
The third and fourth authors assisted with the data parsing and analysis of the institutional data sets. The third author also participated in designing the PD workshops. The fifth author oversaw the data scraping and the sixth oversaw this research as faculty and supported the first author in becoming a full-time member of HJL. 

%Feminist standpoint theory emphasizes seeing from partial perspectives \cite{haraway1988situated}. 
%Through becoming full-time members of HJL, we began to learn what data means to this organization and share their partialities. 
%This will be further elaborated in the findings.
%  or their partiality to data. %particularly the emphasis on qualitative data which acts as a witness and testament to the injustices tenants face.% 
% Becoming a full time member created trust for HJL to share their partialities of data, 
%This situated knowledge is crucial when considering how to design visualizations for a particular use case like community organizing around housing. In what proceeds, we describe our design process and visualizations produced as we transitioned from data intermediaries to accomplices in the fight for housing as a human right, in solidarity with HJL.

\subsection{Evictions and Court Data}
We cannot discuss evictions in a U.S. context without acknowledging histories of systemic racism and capitalism in housing \cite{fields2021racialized,teresa2021eviction,ogbonnaya2020critical}. Evictions engender significant harm, and studies have surfaced how evictions result in negative physical and mental health outcomes and risks of becoming unhoused \cite{desmond2015eviction,clark2019infants}. Studies also point to how this is exacerbated in low-income housing neighborhoods and markets where Black and Brown communities are overrepresented \cite{raymond2021special,raymond2016corporate}. Other studies have pointed to how evictions are not merely a punitive action but a market mechanism in order to restrict the lower housing submarket and further extract financial capital \cite{garboden2019serial,teresa2021eviction}.

Evictions and the data they generate instantiate the standpoint of a juridical system that favors landlords, especially in our geographic context of Atlanta and the state of Georgia.
The very creation of this record is harmful since eviction records severely impinge tenants' credit scores\footnote{In the USA, a credit score is a number between 300 to 850 used by landlords, creditors and employers to evaluate one's financial "trustworthiness"} and deems them a high risk during the tenant screening process \cite{so2022information}.
Therefore, the court data contains information gathered based on a legal system that facilitates eviction.
This data does not consider the standpoint of the evicted.
Additionally, eviction records contain ambiguities and inaccuracies, creating issues in comparative analysis  \cite{PortonDesmond,nelson2021evictions}. 
Counties have different systems and reporting schemas, making generalizations difficult.
As part of this institutional data-gathering effort, the records scraped from each county were standardized into a shared schema to comprise the court data.

Eviction records are generated whenever a landlord files a dispossessory affidavit\footnote{The initial legal proceeding where landlords declare their claim of the property} with the court system to remove a tenant.
%The reasoning can vary, from non-payment of rent to a tenant holding over \footnote{staying beyond the rental contract terms}.
Once this initial affidavit is submitted, a record appears in the digital civic court portal.
Scraping this portal created the court data.
%Subsequent court events for the same case are appended as additional records with the same case ID\footnote{Court events include servicing of warrants, notifications of hearings, submission of evidence and more}.
In this data set, each data item included 26 variables: dates, judgments, addresses, party names (plaintiff/defendant), and others. 
The data was in a \textit{.csv} format, and a single eviction can appear multiple times because each row represented a new event or status update in an eviction record\footnote{Court events include servicing of warrants, notifications of hearings, submission of evidence, and more}.
Before we engaged with the court data, we requested IRB approval and regularly checked in with the data scraping team on how to best handle the sensitive data.

\begin{figure*}[h]
    \centering 
    \includegraphics[width=\linewidth]{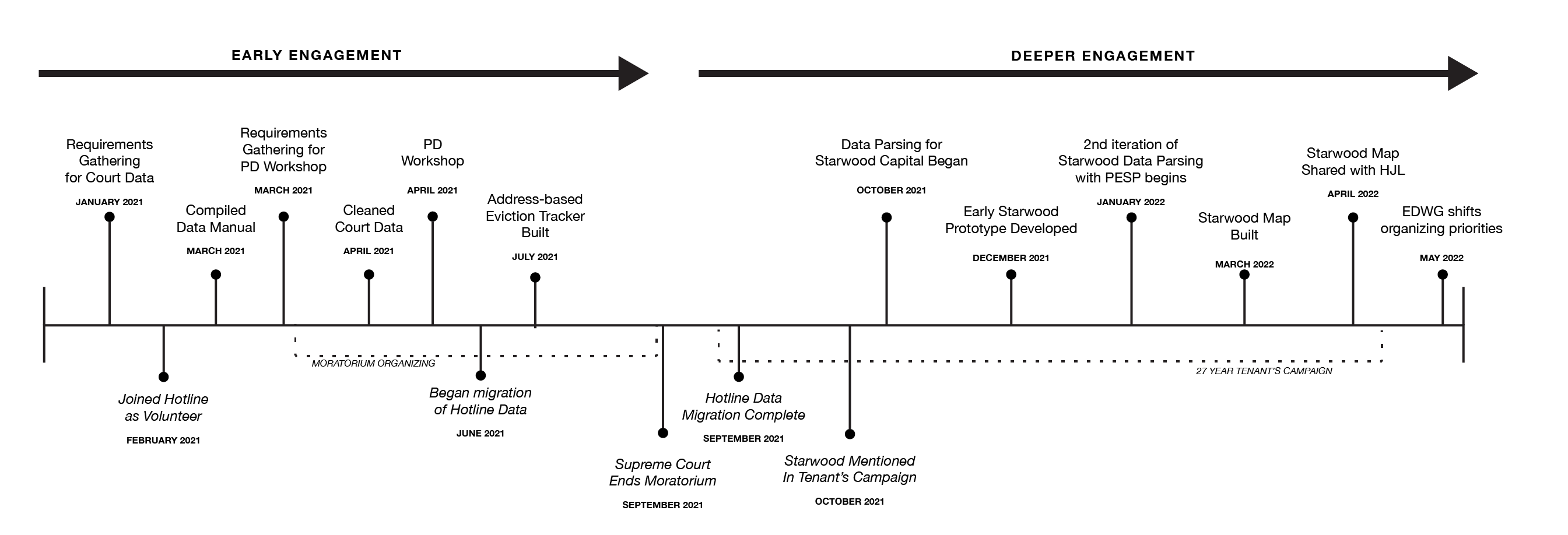}
    \caption{Timeline of the projects.}
    \label{fig:timeline}
    \Description[A timeline infographic]{A timeline infographic that shows points marked by various research activities and activities of HJL.}
\end{figure*}

\section{Methods}
Our research combines ethnography and design, drawing elements from participatory action research and design research to reflect upon the many projects with HJL and conversations amongst ourselves. We developed our paper through an iterative process of reflexive ethnography, paralleling other HCI work in this context \cite{halperin2023temporal,asad2015illegitimate}. For this work, the first author attended 150+ weekly organizing meetings with HJL and participated in direct actions such as protests, canvassing, and field visits. We used Data Feminism to inform our engagement from the onset, detailing how we worked \textit{with} our partners to analyze and understand the local eviction landscape and consider actions we can take, academic or otherwise. This echoes calls within Participatory Action Research (PAR) on structuring such engagements \cite{wadsworth1993participatory,hayes2011relationship}. Our capacity to bring in Data Feminism changed over time due to how our positionality shifted, which we detail in this paper. Our findings are organized between two stretches (Early; Deeper) defined by our positionality at the moment. These positionalities influenced what methods we employed alongside the Data Feminist principles we sought to orient our work around.

We began this community collaboration as data intermediaries, primarily using more standard participatory design approaches and engaging in participant observation as a collaborator.
As data intermediaries, we attended HJL meetings as technical support, trying to understand how they wanted to leverage the court data.
We had one-on-one meetings with working group members to understand needs, which involved observing user walkthroughs of existing data tools and projects. We also conducted semi-structured interviews with the institutional team (6 total) behind the scraping effort to leverage the court data better.
Towards the end of this initial period, we conducted a participatory design workshop with the eviction defense working group that scaffolded data goals and design insights on what data artifacts HJL needed.
Coupled with this workshop was an incomplete visualization dashboard of the court data built by previous students who worked with HJL through a housing course.
We used the visualization design study methodology to identify user tasks, combining them with our partial understanding of the court data to complete and enhance these visualizations (Address-based Evictions Tracker, Fig. \ref{fig:vis1}) detailed in our findings \cite{sedlmair2012design}.

At the conclusion of the first stretch of work, there was no clear research output or novel use of the court data. However, the first author stayed with HJL as a member organizer. This  entailed working shifts on the Tenant Power Hotline (HJL's eviction mutual aid), creating protest graphics, and canvassing. Eventually the first author became a steward rather than just a volunteer for HJL efforts like the hotline. This involved not just working shifts but helping to train volunteers and organize HJL's community collected data.
Throughout this time, if expertise around working with data came up, the first author would loop back with the research team to discuss if there was a way to support these tasks.
This was when the research team built a second visualization, the Starwood Map (Fig. \ref{vis2}), to help HJL identify a specific landlord for an ongoing organizing campaign.
We initially did not consider the organizing activities as research.

The second Starwood Map led to significant changes to HJL organizing priorities and strategies. Deeply embedded in the organization at this point, we sought to understand the difference in outcomes between the two visualizations. 
We turned to research prioritizing first person methods to inform our iterative reflections on our work \cite{howell2021cracks,anokwa2009stories,costanza2020design, halperin2023temporal,asad2015illegitimate}. HJL documents each organizing meeting through a Google Doc created annually. Notes are written ethnographically (Name: Transcript). These notes are shared publicly with the broader organization and those interested in joining. We analyzed 283 pages of meeting minutes across a period of 16 months. In addition to these meeting notes are the research team's personal field notes when conducting visits and interviews. The first author also collected voice memos after each field visit and auto-transcribed them. In this paper, we report the findings and discussions that emerged from revisiting these notes retrospectively, and using Data Feminist principles to reflect upon and evaluate our design process with HJL. Specifically, we interrogated how two different design projects resulted in different outcomes and impacts for the organization.

\section{Early Engagement} \label{placeholder}
In this section, we describe how we worked with the court data and eventually situated it to be effective for our partners. 
In line with the principles of \textit{Data Feminism}, describing these activities aims to \textit{make the labor of opening a public data set visible} \cite{d2020data}.
We begin by describing the initial period of requirements gathering and understanding the context of the court data \cite{d2020data}, which led to the development of the first visualization.
The timeline of our activities is summarized in Figure \ref{fig:timeline}.

\subsection{Contextualizing the Court Data}
\textit{Data Feminism} implores us to always \textit{consider the context} of data \cite{d2020data}. In this section, we describe how the first arc of our work with HJL was oriented around this principle.  
As data intermediaries, we had access to the complete court data set, including sensitive fields removed for the public\footnote{De-identified data that provided aggregate eviction filing numbers were provided on the institutional website, but the address-based information could only be accessed through the complete data set.}.
The complete data set was provided in \textit{.csv} format for each of the five major counties of our metropolitan area.
While this format was accessible to us, we knew it was challenging to interpret for HJL.
Since this data also contained defendant names (those evicted), our institution did not want the complete data set shared. 
These initial constraints were why we were brought on as DIs. Our goal was to find ways to put this data to use for our partners while protecting the sensitive fields.
Before we could consider uses of the court data, we had to understand its context.

We interviewed legal professionals and the institutional data architects to understand the provenance of each field and what they meant. 
We learned that once scraped, the court data was standardized to make it accessible to policymakers, legal experts, and academics.
Each field required specific expertise for it to be of any use. 
For example, judgments and case events are primarily legible to legal aid professionals since each county may have different naming conventions and case-handling events (Table \ref{table:dataTable}). 
In other words, this court data reflected the various standpoints of these professional stakeholders who must convey a sense of objectivity.
However, this data was not immediately accessible to community organizers.
We learned that HJL was only interested in the occurrence of an eviction filing, not all steps of the legal proceeding.
For HJL, an eviction is an act of violence. Counting that act was more important than the legal back and forth.

Our initial cleaning of the court data filtered out sensitive information.
Knowing someone’s eviction history is personal and has a history of use as a means to systemically discriminate against individuals seeking housing, particularly Black communities and People of Color \cite{so2022information}.
Therefore, our institution did not want to make all of the court data public, removing specific fields like the defendant's name. 
Since we were working from the standpoint of DIs, we complied with this request.
To restore context in order to make this knowledge accessible to HJL, we compiled a Data Manual to explain how this data was produced and what each variable meant \cite{d2020data} (Table \ref{table:dataTable}). Beyond the table, we provided annotated screenshots of the digital portal to mark the elements the web scraper collected off the web page, and a systems diagram of the various actors and systems the data moved through.

\begin{table*}
\begin{tabular}{| p{0.2\linewidth} | p{0.55\linewidth} |}
\hline \textbf{Variable} & \textbf{Description} \\ \hline

fileDate & The date this record is created. \\ \hline 
caseID & The unique ID made of numbers and letters. Each county has a different syntax for these IDs. \\ \hline 
plaintiff & The person who is filing for the eviction. It is not always the landlord but LLC companies or agents for multi family unit evictions and serial evictors. Often, single family unit eviction filings will be the landlord. \\ \hline 
plaintiffAddress & The address of the plaintiff, which is often a business address. \\ \hline 
plaintiffCity & The city of the address of the plaintiff. \\ \hline 
plaintiffPhone & The plaintiff’s phone number if it is included as part of the address entered. This is uncommon and only happens in Cobb county so far. \\ \hline 
plaintiffAttorney & The attorney for the Plaintiff. There can be more than one. \\ \hline 
defendantName1 & The tenant (often whose name is on the lease) who is being evicted. \\ \hline 
defendantAddress & The tenant’s address. This will tell you what apartment complex or house the eviction is happening at. \\ \hline 
defendantCity1 & The city of the tenant’s address. \\ \hline 
defendantPhone1 & The tenant’s phone number if it is recorded in the address. \\ \hline 
defendantAttorney1 & The tenant’s attorney. Rare for a tenant to have an attorney. Often you will see “Pro Se” which means they are representing themselves. \\ \hline 
defendantName2 & Other defendants. Usually this is entered as “all others” which means the eviction is including anyone else that lives at that address. \\ \hline 
defendantAddress2 & Usually the same as the main tenant’s address. \\ \hline 
defendantCity2 & The city of the tenant’s address. \\ \hline 
defendantPhone2 & The tenant’s phone number if it is recorded in the address. \\ \hline 
defendantAttorney2 & If the second defendant had a different attorney but usually not the case. \\ \hline 
caseStatus & The status of the case. Usually “OPEN” or “CLOSED” but each county has different names and values. For instance, Dekalb also has a status of “Administratively Closed” along with “OPEN” and “CLOSED.” \\ \hline 
eventNumber & The number order of when a court event happened. (i.e. the first event has the eventNumber of “1”) \\ \hline 
eventDate & The date of the court event. \\ \hline 
eventName & The name of the court event. \\ \hline 
eventDescription & We have not seen this collected even though there is a field for it. \\ \hline 
judgementType & The nature of the case. Usually lists a value like “dismissed” or “order and judgement” meaning there was a court order and judgement because the defendant went through many formal steps of fighting the eviction. \\ \hline 
judgementFor & Who won the case, the Plaintiff or Defendant. This is not always listed, even if a case has a status of “CLOSED” \\ \hline 
judgementComp & The amount of money awarded to the winning party. It is unclear if this includes fees. \\ \hline 
address & The address that the event is about. This is usually the same as the defendant/tenant’s address. \\ \hline

\end{tabular}
\caption{Data variable table of the court data included in the Data Manual created for HJL.}
\Description[A table of data variables.]{A table of data variables of the court data}
\label{table:dataTable}
\end{table*}

After gathering requirements, we conducted a PD workshop with HJL's Eviction Defense Working Group (EDWG), which operates the Tenant Power Hotline. The hotline provided information and support for tenants facing eviction as a bridge toward building tenant organizers. We learned that HJL collects grassroots eviction data through the hotline. This data was a case record of a tenant's housing situation. We envisioned the workshop as a way to scaffold and build data capacities with HJL to make the court data useful to their goals. Based on feedback from HJL during the workshop's design process, we also incorporated their hotline data, working to synthesize how these datasets could be combined. The workshop aimed to identify what questions HJL wanted to answer with data that aligned with their goals. Ideally, identifying these questions would point to potential uses of the court data that could be coupled with their hotline data, or it would point to other data sets that we could eventually scrape. 

\begin{figure*}
    \centering 
    \includegraphics[scale=0.4]{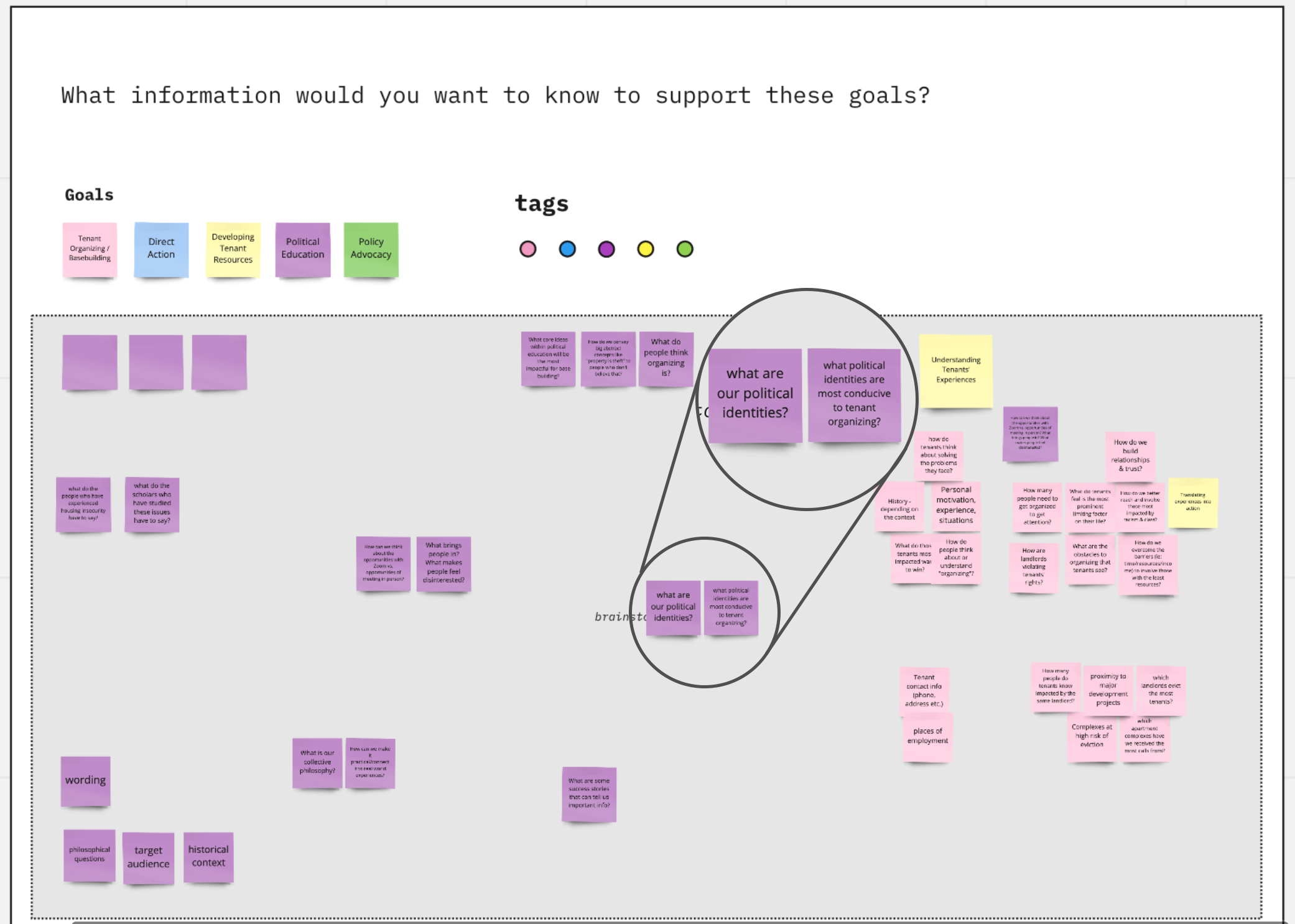}
    \caption{An overview screenshot of the second workshop activity where participants brainstormed questions for affinity mapping. Two questions are magnified in this screenshot: (1) "What are our political identities?" and (2) "What political identities are most conducive to tenant organizing?"}
    \label{fig:Activity2}
    \Description[A screenshot of a digital whiteboard]{A screenshot of a digital whiteboard depicting a workshop activity}
\end{figure*}

We conducted the 3.5 hour workshop virtually through Miro. We designed a post-it brainstorming exercise where members would write down questions they wanted to ask to achieve specific goals. The idea was to take these questions and break them down into "queries" or questions that data could answer. Instead, we received numerous questions about political values (Fig. \ref{fig:Activity2}).
Many of the questions were normative statements that could only be deconstructed into measurable queries through deep discussions.
However, the workshop was not long enough to support these discussions at length, and ultimately, no new uses for the court data surfaced.
%Instead, the workshop challenged our positivist assumptions about what role data can play and the values we saw in data.

Aside from the workshop, there was an ongoing effort to distribute eviction manuals as a free resource at HJL. 
Knowing eviction locations was essential to determine where to deploy the manuals.
In an early meeting prior to the workshop, we observed members trying to find drop off sites based on a dashboard of the court data produced by previous students using Tableau.
Due to the size and nature of the data, which required pre-processing, the Tableau visualizations had severe performance and loading issues. 
We observed HJL members attempt to lookup zip codes with high evictions to identify manual drop-off sites, but this project was never completed because of the frustration and inability to quickly zoom, filter, and find data in the Tableau maps.
Since the PD workshop did not surface new uses for the court data, we focused our subsequent efforts on supporting this existing task of identifying manual drop-off sites with a visualization dashboard. 
We detail the design of this artifact in the following section.

\subsection{Address-based Evictions Tracker}

HJL had early visualization prototypes of the Address-based Evictions Tracker. 
However, in order to determine where to deploy the eviction manuals, they wanted an updated version to support additional features such as rapid dynamic queries, interactive exploration, zooming, filtering, and searching of the data set.
Furthermore, we learned that the initial prototype was hosted on the previous student volunteers' Tableau account.
This led to an uneasy situation where HJL could not own or maintain the visualizations.
Therefore, we decided to recreate and host this dashboard in a shared code repository like Github, where HJL could manage the data themselves if our collaboration ended \cite{hayes2011relationship}.

\begin{figure*}[ht] 
  \centering
  \includegraphics[width=0.98\linewidth]{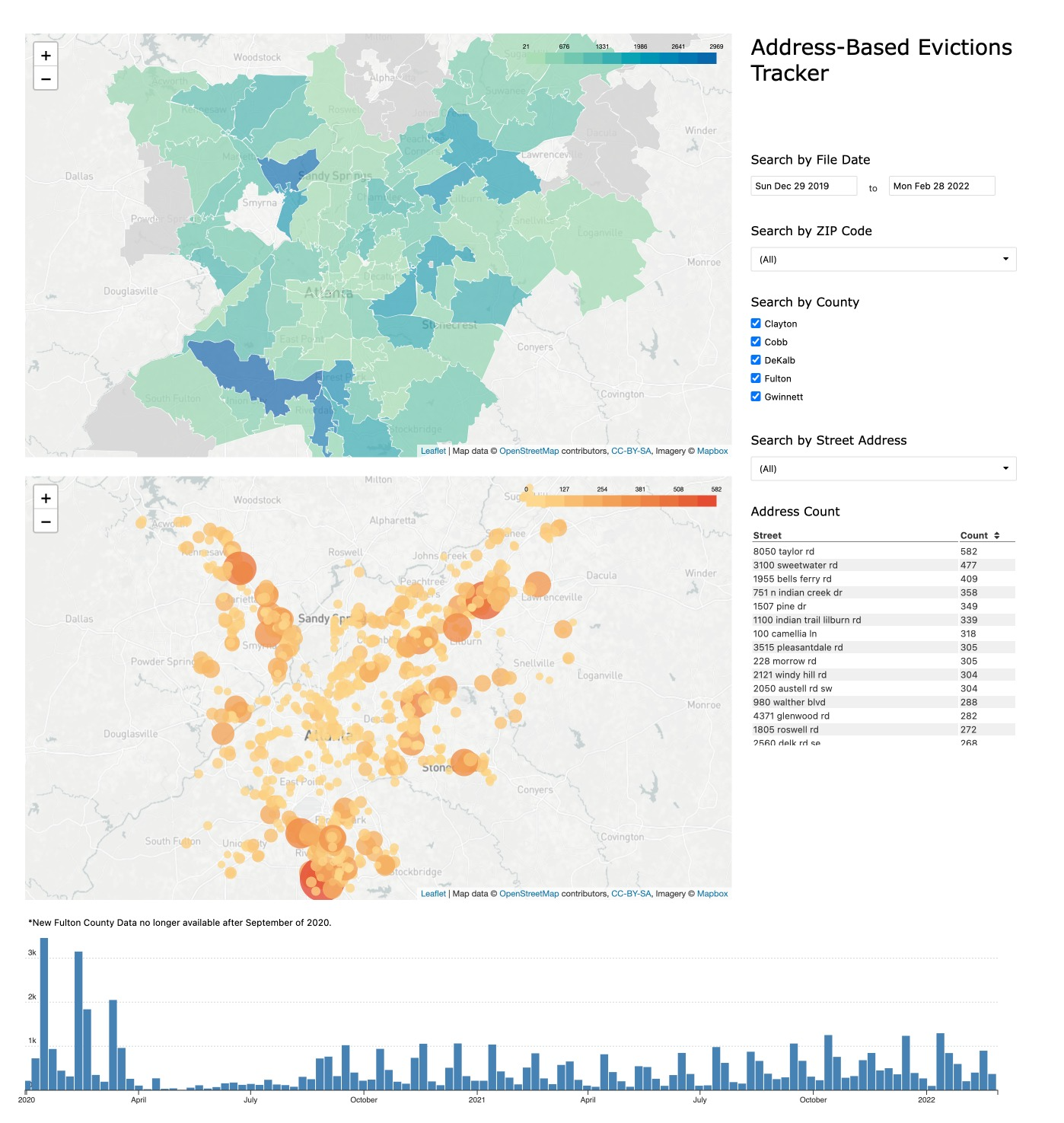}
  \caption{The Address-based Evictions Tracker consists of three visualizations and a control panel. From top to bottom: 1) A map of evictions in each ZIP Code within Atlanta. 2) A map of evictions at addresses within Atlanta. Each point represents an address where one or more evictions took place. 3) A histogram visualizing the total count of evictions in Atlanta over time. Users can zoom in on both maps to see the data in greater detail. To the right of the visualizations is a control panel that allows users to filter by date, ZIP Code, County and Street Address. Clicking on an area or a point displays a tooltip with information about that particular location and a histogram of past evictions there.}
  \label{fig:vis1}
  \Description[A visualization dashboard.]{A visualization dashboard that consists of two maps and a histogram.}
\end{figure*}

The Address-based Evictions Tracker dashboard has two main visualizations (Fig. \ref{fig:vis1}).
Both maps are centered by default on the greater Atlanta area, and users can pan or zoom in to view specific regions in greater detail.
In line with existing visualization best practices, we implemented dynamic filtering \cite{shneiderman1994dynamic} so both maps and the table of address counts automatically respond as filters are applied.
This dashboard was periodically updated with new entries from the court data until the public scraping effort stopped at the end of Summer 2022.
At the last update, there were 346,635 entries.

We presented this Address-based Evictions Tracker dashboard to HJL at a weekly meeting, where one member suggested using the visualization dashboard to identify landlords of addresses with the highest eviction rates.
This desire reflects HJL's wish to \textit{examine and challenge power} \cite{d2020data} using the court data.
However, we quickly realized that the dashboard could not support those tasks.
While the visualizations were able to highlight the relevant addresses, the court data was pre-processed to exclude ownership information for these properties.
All defendant and plaintiff name variables had been de-identified in the dashboard to protect the identities of people evicted, and, in the case of the plaintiffs, because they often filed as LLCs or shell companies, the data was noisy \cite{travis2019organization,raymond2016corporate}.
Taken together, this meant that identifying the landlords using the Address-based Evictions Tracker required HJL volunteers to search addresses on external portals manually.
So while the dashboard made the data accessible and legible for HJL, it could neither \textit{examine nor challenge the power} \cite{d2020data} imbalances in housing.

\section{Deeper Engagement}

In this section, we detail our involvement with HJL after the PD Workshop and constructing the Address-based Eviction Tracker. 
After the workshop, the first author stayed on with HJL as a member and steward. 
We knew HJL wanted to identify landlords with the court data. Furthermore, organizing priorities also began to shift during this time. To support these changing goals, our work became oriented around the Data Feminist principles of \textit{examining and challenging power} and \textit{embracing plurality} with data sets to meet their needs \cite{d2020data}. This resulted in a second visualization, the Starwood Map, that was successful at connecting to the bodies and emotions represented in the data to a neighborhood where HJL was organizing. This period reflected a shift in our positionality from intermediaries to accomplices.

\subsection{Different Data Values and Shifting Evictions Landscape}

Eventually, the first author grew in their role at HJL, from working the hotline to helping manage it. 
During calls, volunteers collect qualitative eviction data through an intake form, capturing information that the court data does not: maintenance issues, illegal actions landlords have taken, and rent burden. 
Collecting this data during shifts was how we gained an understanding of what data means to the organization. 
A working group member proposed moving the community eviction data gathered from the hotline to a new platform. 
The first author volunteered to steward this process and, through this project, became more knowledgeable of the hotline data and the related data practices.
Moving HJL’s data was important in the context of tenant organizing in 2021. 
During trainings for the hotline, we noticed how HJL framed the work as \textit{emotional labor} \cite{d2020data}.
The hotline data is a form of relationship building.
It does not flatten the tenant experience to standardized answer choices; the data reflects this in its messy, qualitative nature. 
Court data could give hard numbers on evictions as they occur and where they occur, but it did not have granular detail about the tenant-landlord relationship in each case. 
After months of volunteering and stewarding their data, we realized how data was not merely a form of counting but a way to witness and preserve testimonies of the injustices of the housing system.

For example, a feature of the low-income housing market is to utilize eviction to assert power and keep tenants restricted to the landlords within that market \cite{teresa2021eviction,garboden2019serial}. This can manifest as serial evicting \cite{garboden2019serial} and generating that eviction record impinges rental applications for new housing during the screening process \cite{so2022information}.
Given this feature of lower housing submarkets, the hotline witnessed severe neglect of properties. 
Callers reported accruals of rent debt coupled with gross negligence of properties, primarily in properties owned by large corporate landlords. 
However, as evidenced by the Address-based Evictions Tracker, the court data alone could not reveal true ownership because landlords can hide behind LLCs and property management companies when listed under "plaintiffName" \cite{travis2019organization,raymond2016corporate}.
Whereas tenant information is legible under "defendantName".
This speaks to the context of the court data (Table \ref{table:dataTable}), which sees from the viewpoint of a legal system that favors landlords over tenants.

Using eviction data to identify eviction hotspot neighborhoods to deploy manuals moved down the priority list.
By the end of the summer of 2021, the U.S. no longer had any federal eviction protections in place. Thus, the aspects of the court data that were useful changed. This speaks to the locality and context of data \cite{loukissas2019all}, which shifted during our time with HJL.
The court data counts in a way that mirrors the asymmetrical power landscape between the tenant and landlord.
\textit{Examining and challenging power} \cite{d2020data} by revealing ownership behind LLCs would require a larger effort.

\subsection{Collating Data Partialities}
In a neighborhood campaign to keep a tenant of 27 years in her home, other residents on her street began to approach HJL about their issues. 
One tenant mentioned that a private equity company, Starwood Capital, now owned their home.  
Identifying Starwood and the other large perpetrators of housing injustice became a new focal point in HJL's organizing strategy.
For the next six months, we endeavored to figure out how to reveal landlords through the data sets we could access.
Describing this work \textit{exposes the labor} of opening the court data, demonstrating how complex the work can be to support a simple task.
We detail this data process to illustrate how we needed to \textit{embrace pluralism} \cite{d2020data}, or multiple data sets and organizations, to get an accurate picture of a single landlord.
Feminist standpoint theory contends that all knowledge is partial and situated \cite{haraway1988situated}.
This is no less true for data.

Limited landlord information in the court data introduced challenges to identifying properties owned by Starwood Capital Group.
When we looked up ownership records for the tenant who mentioned the firm, the name was ``SFR ATL OWNER IV"\footnote{This information is public record via the Tax Assessor and Secretary of State Corporations Division}. 
At this point, we only had our court data and this name. 
Additionally, the court data came from two periods of data scraping. 
Thus, we needed to concatenate multiple data sets across different temporal ranges and counties, resulting in record duplication and completeness issues.
We performed a database join, matching and eliminating duplicates using the unique case ID associated with each record.
We used this court data for our initial attempt to identify Starwood Capital properties by tokenizing the string "SFR" and using it to find matches for \emph{plaintiff}.
We took the paired \emph{defendantAddress} to identify a property (Table  \ref{table:dataTable}). 
We then filtered duplicate addresses, a signal of serial evictions, to populate a map of suspected Starwood-owned properties.
However, because this initial attempt utilized only court data, it provided an inaccurate picture of Starwood's property holdings. 

Our subsequent iteration surfaced after engaging with an advocacy group, Private Equity Stakeholder Project (PESP). 
PESP works with communities negatively impacted by private equity investment.  
They collect data to keep track of these negative impacts and reached out to HJL to see if their data on private equity's housing holdings could be helpful. 
Because Starwood Capital Group is a private equity company, we looked at PESP's data procurement and processing techniques.
We learned that PESP gathered its data from two sources. 
To get the properties owned by a firm, they utilized LexisNexis\footnote{https://www.lexisnexis.com/en-us/gateway.page}, a data brokerage with lists of property tax assessment records. 
To get a count of evictions, PESP utilized re:SearchGA\footnote{https://researchga.tylerhost.net/CourtRecordsSearch/Home\#!/home}, a pay-to-use data portal created by the company that creates enterprise software for court record management.
PESP could identify names of shell companies from LexisNexis and use those names to find evictions filed in re:SearchGA.
However, they could not know the address of an eviction since re:SearchGA omitted address data in the filing.
This speaks to the partiality of each data set.

We also learned from PESP that "SFR" corresponds to "single-family rental," which identified the flaw in our initial attempt to locate Starwood. 
PESP provided a list of shell company name permutations they knew Starwood used.
They also queried a list of Starwood Capital-owned properties for us through LexisNexis. 
These datasets allowed us to cross-reference the court data, which had address information.
We first utilized PESP's list of shell company names directly by matching the \emph{plaintiff} field via an exact string match. 
However, the result only yielded one exact match, suggesting the need for manual pattern examination. 
We learned of additional name permutations and had to parse the Secretary of State's (SOS) Corporation database to complete PESP's list. 
We manually entered each known Starwood name into the Georgia Corporations Division search tool to uncover other permutations that PESP's list may not have encompassed. 
The patterns extracted from this search, in the form of regular expressions, were organized into a table. 
Any plaintiff string could be deemed a Starwood company name by applying each pattern from the table to the string and seeing if there was a match.

Our final step in this process entailed collating Starwood-filtered court eviction records and PESP's list of known Starwood properties from LexisNexis. 
Since PESP's data differed from the court data, we were interested in merging these two sources to fill in gaps.
This resulted in the most comprehensive list of known Starwood properties we could collate.
It ultimately took two organizations and four data sets (court data, re:SearchGA, LexisNexis, and SoS Corporations) to find these properties.
In the next section, we describe how this work culminated in a visualization that exposed Starwood Capital property holdings in the area.

\subsection{The Starwood Map}

\begin{figure*}[h]
  \centering
  \includegraphics[width=\linewidth]{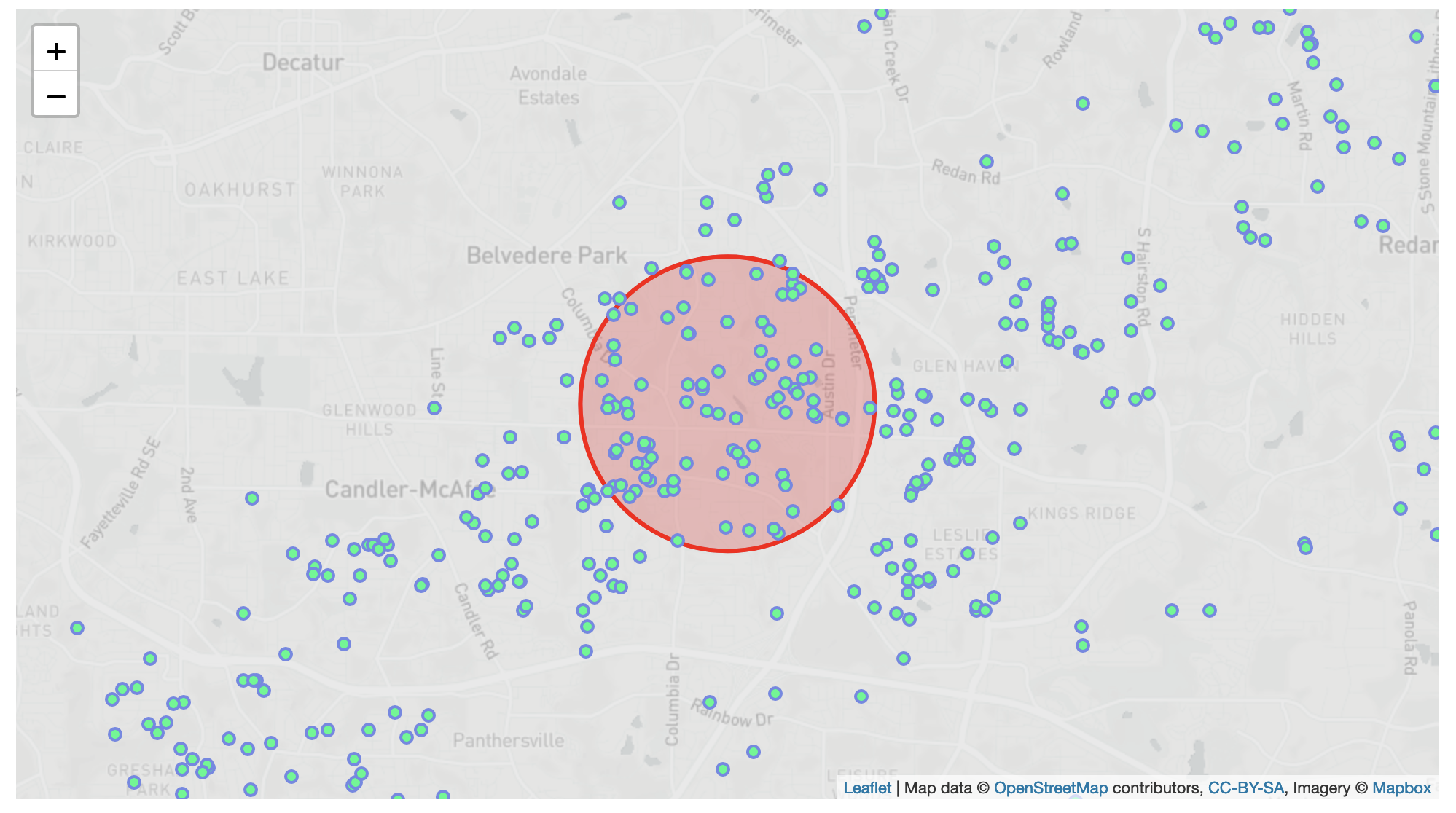}
  \caption{The Starwood Map. A map visualizing the locations of all Starwood Capital Properties in Atlanta. On load, the map is zoomed in on a 1-mile red circle centered on the address of a tenant member of HJL. Users can zoom in further for a more detailed view or zoom out for an overview of all Starwood Capital Properties evictions in the greater Atlanta area. Clicking on a point also brought up a tooltip with the specific address and the data provenance.}
  \label{vis2}
  \Description[A single map visualization.]{A single map visualization that displays dots on a map.}
\end{figure*}

The resulting Starwood Map (Fig. \ref{vis2}) consists of a single map visualization of the 1014 properties owned by Starwood Capital that we could identify.
A single point represents a property address.
We also visualized a 1-mile radius centered on the address of a tenant member of HJL behind an ongoing neighborhood campaign (red circle, Fig. \ref{vis2}). At this point, HJL had worked with this tenant for nearly a year, produced a crowdfunding campaign, and engaged in multiple direct actions against the landlord and their associates. Additionally, organizers had spent months canvassing the neighborhood when the name "Starwood" got on HJL's radar in the first place.

This Starwood Map differed vastly from the Address-based Evictions Tracker in terms of what data was included in the visualization.
It used a far smaller data set, including only evictions instigated by a single landlord, Starwood Capital Properties.
However, when we presented this visualization at an annual retreat for the Eviction Defense working group of HJL, we were able to observe how it affected concrete changes in HJL's organizing strategies.
By centering on and highlighting the area around an ongoing neighborhood campaign, the visualization confirmed community accounts of what was happening and what HJL suspected was happening across the city through the hotline. 
It emphasized the scale of ownership PE firms held in neighborhoods HJL canvassed.
The red radius was also intimately familiar to HJL members, who had built \textit{emotional ties with people} living there \cite{d2020data}.

After the retreat, the working group voted on shifting their organizing practices.
In the past, the hotline and the data HJL had access to were used in a reactionary manner. 
If someone faced a crisis or an issue, HJL would react and shift attention to those needs. 
This led to overcapacity issues. 
After the Starwood Map presentation, HJL elected to use the data proactively. 
Instead of answering where they were called, HJL would focus their organizing on targeting corporate landlords where they had relationships with tenants.

\section{Discussion}

So far, we have recounted our work applying Data Feminism to situate a large, public data set for grassroots tenant organizing. We detailed how we collected, organized, analyzed, explained, and visualized the court data to HJL. 
Data Feminism principles tell us \textbf{what} we should strive for but not necessarily \textbf{how} \cite{d2020data}.
Reflecting on our experiences working with the court data to design the Address-based Evictions Tracker and the Starwood Map, we describe the challenges of putting principles into practice to rethink HCI design conventions.
In this section we describe three implications for design in grassroots contexts that begin to move from what to how: becoming a domain beginner, striving for data actionability, and evaluating our design artifacts by the social relations they sustain rather than just their technical efficacy.
These implications may be relevant for other HCI scholars when structuring their engagement with community partners.

\subsection{Becoming a Domain Novice to Become a Data Accomplice}
%Sign post paragraph

Shifting our positionality from an intermediary to an accomplice was necessary to situate the eviction data set. 
%What was vital in this process was to develop a hybrid subjectivity that allowed us to really consider and understand the context \cite{d2020data}. 
%In data visualization, robust methodologies \cite{sedlmair2012design, mccurdy2016action, syeda2020design, mckenna2014design} guide how visualization researchers engage with specific users and data domains.
%Implicit in these study frameworks is a division of labor between the domain expert and the technical expert.
While we were experts in data, visualization, and analysis, we had no expertise when it came to tenant organizing.
Shifting our positionality required us to \textbf{embrace a positionality of a domain novice}, rather than compartmentalizing ourselves from that part of the work altogether.

We began this project relying on approaches based on our understandings of data and our skill sets in PD, Data Science, and Visualization. In our role as data intermediaries, we took-for-granted a clear division between the domain expert and the technical expert. This was also in line with common visualization practices that consider researchers and domain experts as separate and distinct entities \cite{sedlmair2012design, mccurdy2016action, syeda2020design, mckenna2014design}. We hoped to offer our technical skills to help HJL leverage the court data. However, the questions participants raised in the PD workshop could not be easily translated into questions one can ask of data. Naively, we thought we could just bring our technical expertise and our partners would know what to request of us after we explained what we could do with the court data. Our approach reiterates Harrington et al.'s call to critically examine how PD workshops fail to account for critical factors when designing to address inequality \cite{harrington2019deconstructing}. Similarly, our experience collaborating with HJL challenges common visualization practices, demonstrating the need to develop our own domain competency to formulate the right visualization tasks. We could not understand the value and needs of data for HJL until we really engaged in the day-to-day activities of the organization. In other words, we needed to set our technical expertise aside to become domain novices.

% Drafting [d]

%As Passi and Jackson note, data systems are the result of a "collaborative accomplishment" involving translating between different ways of knowing \cite{passi2018trust}.

%To gain domain competency, one must engage in the mundane tasks of an organization as a novice. necessary for a collaboration. ... Best practices will depend on the domain one is working in. For instance, Passi and Jacksons's work in corporate.

%Passi and jackson emphasize the immportance of translation in dat acience between different wasy of knowing and out initial owrkshop asked participatns to do the translation work for us. However to be in mutual conversation required us to have both domain expertise and technical experits. Similar issues have been raised by Sum et al.... 

Our work builds on Passi and Jackson's study examining data systems as the result of complex collaboration \cite{passi2018trust}, but in the resource constrained context of tenant organizing.
%How this collaboration came together in our context was different given that it is grassroots-based versus a formal corporate setting.
The role of the person bringing technical expertise as an outside intermediary requires developing a domain competency, which means one must embrace coming in as a novice. For us, this involved attending weekly eviction work meetings, working shifts on the hotline, or door-knocking in neighborhoods where campaigns occurred. Only through routinely engaging in these activities could we truly understand how to find clear usage amongst shifting priorities. It was the mundane tasks that helped us achieve this standpoint, rather than the more intellectually-attuned work of devising strategies and speculation. This was evident in how our PD workshop failed at producing any tangible takeaway. Alongside this point is an inherent risk when researchers begin the terms of an engagement leading with their proficiencies and capacities. In retrospect, we saw how positivist values were brought into the workshop. The simple exercise of turning normative statements into something query-able fell apart with the critical questions HJL wanted to ask and answer with data (Fig. \ref{fig:Activity2}).
This reflected a gap in our domain understanding, showing how we had yet to truly engage in organizing at that point.

%Standpoints are earned
Stepping into organizing was a means to truly gain context in our research and structure that into the terms of design participation and engagement that Data Feminism advocates \cite{d2020data}.
By stepping in, we could make sense of the mess of shifting priorities in a changing eviction landscape. 
This in turn allowed us to grapple with heterogeneity and partiality of the various data sets needed to produce a useful artifact \cite{passi2018trust}.
This journey to become accomplices when we began as intermediaries reflects the political process standpoint theorists describe \cite{collins1986learning,harding2004feminist,harding1986science}.
Liang et al. discuss the tension of allyship in HCI research, describing specific points for how scholars can center the agenda of marginalized groups\cite{liang2021embracing}.
We had to act as accomplices for HJL and the fight for housing justice to truly understand their situated knowledges.
In this way, our work politically engages from below and takes an explicit activist orientation \cite{whitney2021hci,asad2015illegitimate,asad2019prefigurative}.

%Accomplice vs ally
We build on these tensions and answer calls within our community to move beyond allyship towards accompanying our partners \cite{asad2019academic}.
Acting as an accomplice does not mean abandoning your expertise, but knowing when to defer your expertise until it is needed. 
It is acknowledging that your expertise has no claim or precedence over your partners'.
Accomplices allow the work to be directed by those oppressed by the structures they are trying to dismantle \cite{clemens2017ally,indigeneousaction_accomplices_2014}. 
Ghoshal et al. identify pitfalls in grassroots collaboration such as when certain technical tools and technologies exclude others from conversations and goes against the values of these organizations \cite{ghoshal2019role,ghoshal2020toward}. \textbf{Becoming a domain novice places the researcher in a position where it becomes clear who should direct the work}, which in turn mitigates many of these pitfalls.

\subsection{From Accessibility to Actionability}

% Role of data in a community engagment
% Prior open data works often crituqing the avaiblibility
% While may be avvaiblae may not be accessible (returning context to the data)
% Even if accessable is not neccesarily actionable (returning context to the task)
    %  Need to align context of data and context of task

%Signpost Paragraph
In this section, we extend the Open Data conversation by arguing that restoring the context for a data set -  making it accessible -  is not enough for it to be truly open and usable. Instead, \textbf{we must make public data sets actionable}. Many scholars and activists of the Open Data Movement caution that just releasing public data and making it available is not enough \cite{paudel2023reimagining,d2020data}. D'Ignazio and Klein point to how low funding and poor infrastructure leads to open data sets becoming zombie data sets -  published without a clear use or purpose \cite{d2020data}. The Open Data Movement outlines how making data available can end up sacrificing the data set's context. Publishing this court data is an instantiation of the broader movement to make public data not just available, but accessible. The institutional scraping effort also knew that the potential value of this data for community organizations would require more support than simply hosting an API. Restoring the context of the court data was a large part of our early engagement with HJL. However, the data did not serve any concrete action until the Starwood Map was designed.

%What does Actionability mean? Task Related
We propose three steps data scientists and scholars should follow to make use of public data towards goals of Openness: (1) availability, (2) accessibility, and (3) actionability. Each step is a prerequisite for the next. First, public data must be made available. This is the current status quo of civic data which Data Feminist scholars critique, arguing that while data is available it is not accessible because it lacks context \cite{d2020data,paudel2023reimagining}. Building accessibility is the second step. Our early work took the available court data and made it accessible by providing necessary initial context and building the Address-based Evictions Tracker to allow organizers to easily peruse the data. However, data scholars must also support design endeavors that support a clear action or task. Making data accessible with context does not inherently beget an action or a task. The Address-based Evictions Tracker went far in making the court data accessible to HJL, but it didn't tell them anything they didn't already know: that evictions are rampant and have higher occurrences in Black and immigrant neighborhoods. Understanding the data is not the same as doing something with it. There is additional labor needed to not only contextualize the data, but gain understanding of necessary actions and tasks that are situated in a particular time or place \cite{suchman1987plans}. Matching the context of the data to the context of the task is precisely what is required in order to make data actionable.

%Availability is a prerequisite to accessibility. In our case, making the data available (i.e. sharing the entire .csv) would have done nothing because there was no context and the data was too large to digest and understand easily. The data manual provided necessary initial context and the Address-based eviction tracker allowed organizers to easily peruse the data. These deliverables made the data accessible but not actionable. Making data actionable means finding a way for the data to support a specific task. While the visualization design methodology focuses on how researchers should translate tasks and data from a domain-specific form into a abstract visualization tasks\cite{sedlmair2012design}, our experiences demonstrate instead that tasks are beholden to the specifities of context \cite{suchman1987plans}. Paradoxically, abstraction does not mean pulling away from the subjectivities of the domain. Suchman argues that actions and tasks are situated in particular time and place. Therefore, to abstract a task we must have an understanding of the contextual factors \cite{suchman1987plans}. Accessibility then is a prerequisite and means towards actionability.

%Outcome Comparison to demonstrate Accessibility vs Actionability
Our second visualization demonstrates achieving actionability. The Starwood Map (Fig. \ref{vis2}) differed vastly from the Address-based Evictions Tracker in terms of what data was included in the visualization, the design choices made, and ultimately, the outcomes engendered. While we imagined the Address-Based Evictions Tracker could support the task of identifying manual drop-off sites, by the time it was built this task was no longer relevant to the organizational time and place. In contrast, the Starwood Map made the data \textit{actionable}, and in doing so, effected concrete changes in HJL's organizing strategies. This was possible because the Starwood Map was more than a presentation of all the court data available. It was sensitive to the immediate priorities of HJL, supporting the specific task of identifying landlords of properties for canvassing.
Furthermore, by visualizing the locations of Starwood Properties and their presence in a neighborhood where HJL was already active, the Starwood Map foregrounded the relationships HJL volunteers had already nurtured with local residents.
In other words, the Starwood Map brought together the specific contextual factors of the data and the current contextual factors of the action HJL wished to undertake. These contextual factors emerged out of a particular time and place \cite{suchman1987plans}. We could only appreciate their significance once our standpoints shifted to data accomplices, allowing us to design a visualization that was better aligned with the experiences and everyday organizing needs of HJL.

%Consider visualizations beyond the scheme of communication
% Public viz more about accessable rather than actionable. The viz isn't the action. 
% Prior work the tool is the action, end unto itself. In this work, the tool supports the community, means to an end.
Moving from availability, to accessibility, to actionability has implications when we consider how to apply this framework for critical visualization design and countermapping. 
For visualization scholars, design methodologies, such as \cite{sedlmair2012design}, often focus on how researchers should translate tasks and data from a domain-specific form into abstract visualization tasks. In contrast, our experiences demonstrate that tasks are beholden to the specificities of context \cite{suchman1987plans}. Paradoxically, abstraction does not mean pulling away from the subjectivities of the domain but to move closer towards it. 
Moreover, for community engaged countermapping, we often see artifacts that err towards goals of public awareness and coalition building.
The Anti-Eviction Mapping Project (AEMP) \cite{maharawal2018anti, halperin2023probing,halperin2023temporal}, for instance, has designed some of the most poignant visualizations that communicate to public audiences the pervasiveness of evictions and stories of displacement.
In contrast to these externally facing tools and maps, our visualizations are internally facing artifacts that supported everyday organizing tasks rather than external public awareness campaigns.
While visualizations for public awareness and coalition building are important, there is a need for more work exploring the role of visualization to support situated actions.
%Implicit in their design is a goal that the visualization is the action, in other words, the map is an end in and of itself.
Actionability requires the technical expert not to look at maps as ends in themselves, but instead to attend to the actions the artifacts supports.

%Conclusion
In detailing the design of our visualizations, we point to new possibilities for HCI scholars for what sorts of tools can be made. Carrera et al. have called on scholars to ask if "resources and labor that could have been spent on productive activities for the communities they serve are instead redirected towards efforts to educate the privileged or justify their claims on the state'' \cite{carrera2023unseen}. Instead of focusing on public-facing visualizations as a means of rhetoric and communication, we can also build tools that internally support the tasks of activist organizations. The way we arrived at these designs was through striving to make the data actionable to a particular audience, rather than just available or even accessible to a general audience. \textbf{Identifying actionability requires gaining intimacy with the context and matching it to the contextual factors of the data.} Actionability then, should be brought into discussions of Open Data and become a design goal when creating artifacts in justice-oriented research.

% Overly simplisitic to think about maitnrenace solely in the relaitonship between things and tools. Demonstrating care.
\subsection{Sustaining the Tool versus the Relationship}

%signpost paragraph
In HCI and Computing, we often focus on tools and technologies because this is the traditional site of our expertise. Hayes implores us to ensure that a technical tool can be maintained in order to sustain positive social changes \cite{hayes2011relationship}. These concerns resurface in prior work on organizing and public data. Carrerra et al. find that across the grassroots organizers they interviewed, there were sustainability concerns of using public facing digital tools for counter mapping \cite{carrera2023unseen}. Paudel and Soden call attention to the issue of open data platforms going stale if they are not constantly updated \cite{paudel2023reimagining}. While maintaining and sustaining a tool is important, this can depend on the context of the technical artifact and its situated use. In our work, maintaining our technical interventions proved fruitless or impossible. Sustaining the Address-based Evictions Tracker did not result in continued or new uses, and the Starwood Map had a limited window of relevancy to HJL. Nonetheless, tools can be a means to social relations, and each tool served their purpose in nurturing our relationship with HJL. \textbf{HCI scholars should approach making as a means of sustaining a relationship rather than building something novel or useful.} Doing so changes how we evaluate the things we make from solely focusing on what an artifact \textit{does} to also looking at the relationships our artifacts can support \cite{frauenberger2019entanglement}.

We maintained the Address-based Eviction Tracker for over a year, not because of it's value to HJL as an artifact but rather as a way of maintaining our relationships.
This maintenance involved regularly updating the data set and coming up with workarounds to avoid increasing HJL's monthly expense of hosting the visualization. However, even while this map was maintained, it was never used towards any organizing action, and eventually the public scraping effort stopped at the end Summer 2022. This ended our ability to update the Address-based Evictions Tracker. Yet reflecting on our collaboration with HJL, we don't view this Address-based Evictions Tracker as a complete loss. Building and maintaining the tool was a way to show good faith in our relationship with HJL. We promised to build it, and we did. We updated the data to keep it operational, which showed HJL that we could be trusted if other data needs came up.

% Ony able to buidl map because of the realtinoahips built by our eariler work on ABT. Movoer, Starwood furthre problematizes the importance of maitaining tools.

% Sustaining the tool dosn't matter
We were only able to build the Starwood Map because of the relationship developed from our early work and the trust HJL had in our technical capacity. However, this artifact also problematizes the importance of maintaining tools.
The Starwood Map was effective in generating an insight that led to a change in organizing strategy (targeting corporate landlords) and helped identify locations for canvassing. However, these situated actions were only relevant in that particular time and place \cite{suchman1987plans}.
In recent visualization work, Akbaba et al. have suggested  developing visualization artifacts from the orientation of ``designing for the graveyard'' and the acceptance that one day the tool will no longer be used \cite{akbaba2023troubling}.
We detail the process of collating data and constructing the Starwood Map as an example of what this ``designing for the graveyard'' might look like.
The data we put together to identify Starwood properties had dependencies that kept it's relevancy limited. It is not uncommon for institutional investors to exchange and acquire property portfolios from each other \cite{ash2023blackstone}. Tax Assessor records and incorporation documents can easily change within a year. Therefore, the data used to support the organizing task of identifying canvassing targets was only relevant for a window of time.

%talk about working through the messiness of the data as another form of commitment/trust building

%Halperin and McElroy describe potential pitfalls and alternatives to what technological relations may look like over time \cite{halperin2023temporal}. They reflectively detail how solidarity is carefully built. Paralleling their work, Our technical intervention to target worst offenders is similarly built from a patchwork of various resources and data. We ensured that the tools had community control and oversight through how it was hosted and how it was built. All of these design choices developed our technological relation to HJL, built off of our personal commitments as volunteer organizers. 
While neither of our tools could be sustained beyond the timespan of the work described, we would argue that sustaining the tool should not be our only goal. \textbf{Instead of evaluating tools by their efficacy or functionality, we should evaluate them by the social relations they engender, how long lasting those relations are, and how that supports the work of the organization we accompany.} Building trust is an imperative when one works with community organizers \cite{carrera2023unseen}. Our artifacts had more value in how they fostered our ongoing relationship with HJL than their technical value and use. We are not arguing for researchers to refrain from building technologies, but instead to deeply reflect on the greater purpose they serve in community engaged partnerships.

\section{Conclusion}

In this paper, we provide an in-depth and reflexive account of accompanying HJL, a tenants rights organization, to make aggregated public eviction records actionable through data visualization. We used Data Feminism to structure our collaboration and design process and reflect on how we put these principles into practice. We show how shifting positionalities from Data Intermediaries to Data Accomplices impacted our design process, our employment of Data Feminist principles, and the impact of our visualizations. Through our work, we offer three implications for design to HCI scholars, especially in grassroots contexts: becoming a domain novice, striving for data actionability, and evaluating our design artifacts by the social relations they sustain rather than just their technical efficacy.

\begin{acks}
We would like to thank Housing Justice League for trusting us to accompany the larger fight for housing as a human right through our data work. Thank you to Alison Johnson, Natalie McLaughlin, Dani Aiello, Foluke Nunn, Josue Acosta, Graham Kelly, Julie Siwicki, and the rest of the Eviction Defense working group. We would also like to thank our reviewers for their thoughtful feedback. Fifth time's the charm! This work was supported in part by the Atlanta Interdisciplinary AI (AIAI) Network, sponsored by the Mellon Foundation, in part by the Russell Sage Foundation, RSF Grant \#2010-28252 Preserving Rental Housing Stability during Disasters: An Examination of Eviction Moratoria during the COVID19 Pandemic, and in part by IBM Research.
\end{acks}

\bibliographystyle{ACM-Reference-Format}
\bibliography{refs}

\end{document}